# Evaluation of CT Scan Usability for Saudi Arabian Users

Saad Aldoihi
Computer and System Engineering.
ENSTA PARISTECH
Palaiseau, France
saad.aldoihi@ensta-paristech.fr

Omar Hammami
*Computer and System Engineering*
ENSTA PARISTECH
Palaiseau, France
hammami@ensta-paristech.fr

*Abstract*— Like consumer electronic products, medical devices are becoming more complicated, with performance doubling every two years. With multiple commands and systems to negotiate, cognitive load can make it difficult for users to execute commands effectively. In the case of medical devices, which use advanced technology and require multidisciplinary inputs for design and development, cognitive workload is a significant factor. As these devices are very expensive and operators require specialized training, effective and economical methods are needed to evaluate the user experience. Heuristic evaluation is an effective method of identifying major usability problems and related issues. This study used heuristic evaluation to assess the usability of a CT scan and associated physical and mental loads for Saudi Arabian users. The findings indicate a gender difference in terms of consistency, flexibility, and document attributes, with a statistically significant gender difference in mental load.

*Keywords*— *Usability, CT scan Heuristic Evaluation, Human-computer Interaction.*

## I. INTRODUCTION

In performing a CT scan, it is important to ensure effective interaction between the radiologist and the CT scan device itself as well as the patient. In particular, the impeded speech command, which directs the patient to take, hold, and release their breath (available in many languages) is a key interaction between operator and machine. This fundamental task involves multiple disciplines, including Human Computer Interaction, Human Computer Design, and Usability. As much of the scan's accuracy depends on how fluently the operator can interact with the machine, designers are under pressure to extend the machine's capabilities and usability. According to the FDA database MAUDE [1], 437 incidences of "User used incorrect product for intended use" were reported in the years 2016 and 2017, 11 of which resulted in death. For that reason, designers have a serious responsibility to ensure devices usability. This is not a straightforward matter like following a cooking recipe but depends on rules that emphasize goals rather than sets of actions [2]. In most tasks, the interactions between operator and device are goal-oriented; once the user completes the desired task successfully and efficiently, the goal can be said to have been met [3]. In the case of a CT scan, the goal is achieved when the radiologist effectively completes successful testing of the patient.

In relation to usability, safety is a critical consideration when assessing the success of a CT scan. One study [4] showed that CT emits as much radiation as 200 chest X-rays, which for the average person is equivalent to more than seven years exposure in a natural setting. The same article reported that children commonly received adult doses of radiation. In her Testimony [5] before The United States House of Representatives Health Committee Subcommittee on Energy and Commerce, Rebecca Smith-Bindman MD stated that the most common type of CT scan emits a level of radiation equivalent to 1500 dental X-rays, and that in some CT scan models, the level is equivalent to 5,000 such X-rays. In most cases, whether on the basis of fact and reason or unfounded speculation, patients express concern about the possible effects of a CT scan on their health.

As it is difficult to determine or evaluate current usability practices within the medical device industry, medical device usability issues need to be publicized, analyzed, and explained [6]. Other industries such as air traffic control and nuclear energy have benefited immensely from human factors and usability practices to eliminate errors and improve safety [7]–[9]. Ultimately, whatever the industry, human factors and usability analyses are safety-driven [10]. To the best of our knowledge, no existing study has applied heuristic evaluation specifically to CT scans. Any related studies have not measured CT scans as a direct product but rather as part of a picture archiving and communication system (PACS) or medical imaging software. To the best of our knowledge, this is the first paper to specifically apply heuristic evaluation to CT scans.

## II. CONSOLIDATED USABILITY ATTRIBUTES

### A. Baseline attributes

Nielsen and Molich [11] proposed a new method for evaluating usability, which they called "heuristics", and heuristic evaluation has since become a popular tool because of its effectiveness and low cost. Following its successful application in evaluating the user interface, heuristic evaluation has since been adopted in other domains [12]. Nielsen [13] introduced ten heuristics that serve as an evaluation guide for practitioners: 1) visibility of system status; 2) match between system and real world; 3) user control and freedom; 4) consistency and standards; 5) error prevention; 6) recognition rather than recall; 7) flexibility and efficiency of use; 8) aesthetic and minimalist design; 9) help for users to recognize, diagnose, and recover from error; and 10) help and documentation. As many authors in the usability field have sought to develop definitions and a holistic approach to usability, it may be inferred that usability cannot be consolidated as a single attribute, and many attempts have been made to compile a list of attributes. Similarly, Makoid [14] noted that there is no single agreed definition of usability; instead, different definitions may incorporate different parameters and attributes. Nevertheless, there is consensus on the importance of usability, and international organizations such as ISO have introduced usability attributes for the purposes of standardization. As noted by Shneiderman and Plaisant [15], standardization accelerates industry adoption

*B. Improved attributes*

Shneiderman and Plaisant [15] postulated that it is extremely difficult to designers to accomplish the final design without being forced to tradeoffs between attributes. In other word, increasing effectiveness of one attribute comes on the expenses of the another attributes. In order to achieve better yield of discovering usability problem, traditional heuristic evaluation has been modified, extended, and improved to suit a specific domain or task [16] . To Ling and Salvendy [16] heuristics evaluation categorized into three approaches: 1) alteration of the evaluation procedure, 2) expansion of the heuristics evaluation procedure, and 3) extending the HE method with a conformance rating scale. In the medical equipment domain, Zhang, et al (2003) developed heuristics evaluation which are extended and modified version of Nielsen [17] and Shneiderman [18]. Zhang et al. [19] combined Nielsen [17] and Shneiderman [18] to constitute an extended and more fitted heuristics to medical devices.

## III. METHOD

*A. Participant*

Careful sampling is a crucial element of successful research. In user research studies, recruiting participants who meet precisely determined criteria can prove very challenging [20]. Cairns and Cox [21] insisted that recruiting people with specialist knowledge is essential to user study success. To ensure scientific integrity, the authors carefully specified a rigorous pre-qualification procedure to eliminate unwanted participants. To qualify as a participant, candidates had to meet three criteria: 1) their primary work involved handling and operating CT scans; 2) their job category was radiology (Radiologist or Technician); and 3) they were currently working in the Saudi public healthcare system. At the time of this study, there were 400 CT scan technicians working in the public sector, and there were about 191 CT scan devices (according to the Radiology Department in the Ministry of Health, November 2017). In total, there were 44 participants, ranging in age from 20 to 49 years (26 male and 18 female). Table 1 shows the participants' demographic data.

TABLE I. PARTICIPANT DEMOGRAPHIC

| Variable | Frequency | N % |
|---|---|---|
| **Gender** | | |
| Male | 26 | 59.10% |
| Female | 18 | 40.90% |
| Total | 44 | 100% |
| **Age** | | |
| 20-29 | 14 | 31.80% |
| 30-39 | 23 | 52.30% |
| 40-49 | 7 | 15.90% |
| Total | 44 | 100% |
| **Level of Education** | | |
| Diploma | 11 | 25.00% |
| Bachelor | 25 | 56.80% |
| Master | 8 | 18.20% |
| Total | 44 | 100% |
| **Years of Experience** | | |
| 0-3 years | 16 | 36.40% |
| 4-7 years | 11 | 25.00% |
| 8-11 years | 9 | 20.50% |
| 12+ | 8 | 18.20% |
| Total | 44 | 100% |
| **Radiology Job Title** | | |
| Consultant Technician | 1 | 2.30% |
| Registrar | 3 | 6.80% |
| Senior Technician Specialist | 9 | 20.50% |
| Technician | 15 | 34.10% |
| Technician Specialist | 16 | 36.40% |
| Total | 44 | 100% |
| **Nationality** | | |
| Filipino | 7 | 15.90% |
| Indian | 1 | 2.30% |
| Pakistani | 2 | 4.50% |
| Saudi | 33 | 75.00% |
| Sudanese | 1 | 2.30% |
| Total | 44 | 100% |

*B. Instrument*

The study instrument comprised two elements, the first of which was a two-part questionnaire. The first part gathered information about participant characteristics including gender, age, educational level, years of experience, job title, and nationality. The six measured variables included the following options: (1) Gender (Male, Female); (2) Age (20–29, 30–39, 40–49); (3) Level of Education (Diploma, Bachelor, Master); (4) Years of Experience (0–3 years, 4–7 years, 8–11 years, 12+ years); (5) Radiology Job Title

(Technician, Technician Specialist, Senior Technician Specialist, Consultant Technician, Registrar); (5) Nationality (Open).

As shown in Fig. 1, the second element comprised two adapted questionnaires. The first section adapted the standard heuristic evaluation approach to identify major usability issues vis-à-vis CT scan, based on Zhang et al.'s [19] approach to heuristic evaluation. As shown in Table 2, heuristic evaluation tends to measure CT scan usability in terms of 14 attributes: (1) Consistency; (2) Visibility; (3) Match; (4) Minimalism; (5) Memory; (6) Feedback; (7) Flexibility; (8) Message; (9) Error; (10) Closure; (11) Undo; (12) Language; (13) Control; (14) Document. Each attribute was measured by items ranked on a five-point Likert scale (0 = *No Problem*, 1 = *Cosmetic*, 2 = *Minor*, 3 = *Major*, 4 = *Usability Catastrophe*).

In a third step, NASA-TLX [22] was used to assess the physical and mental loads associated with operating a CT scan. The questionnaire measured the following variables: (1) Physical, (2) Mental, (3) Frustration, (4) Discomfort. Responses were again based on a five-point Likert scale (1 = *Low*, 2 = *Fairly Low*, 3 = *Medium*, 4 = *Fairly High*, 5 = *High*).

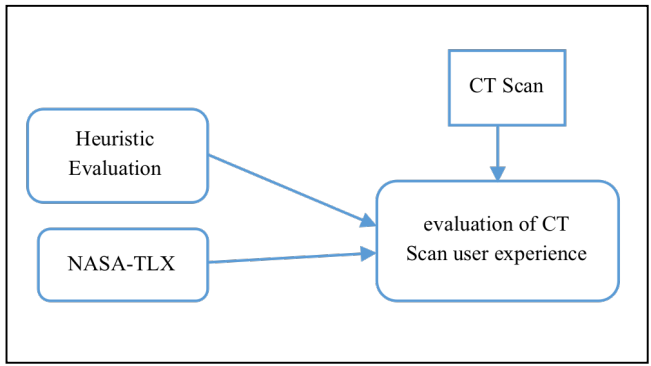

Fig. 1. Model Framework

The second set was field observation. The first author conducted 16 hours of on-site observations at King Saud Medical City (KSMC) to investigate technicians' use of the CT scan in terms of usability and physical and mental loads. During that time, 8 hours were dedicated to CT scan operation in the Emergency Room (ER), and the other 8 hours were dedicated to CT scan operation in the Radiology Department. KSMC was chosen as the observation site after careful evaluation of several Riyadh hospitals in terms of throughput and diversity of patients.

The observed technicians were made aware of the study's purpose, and a consent form was distributed and obtained from each participant.

TABLE II. USABILITY ATTRIBUTES AS DEFINED BY ZHANG [19]

| Attribute | Explanation |
| --- | --- |
| Consistency | Consistency and standards: Users should not have to wonder whether different words, situations, or actions mean the same thing. Standards and conventions in product design should be followed. |
| Visibility | Visibility of system state: Users should be informed about what is going on with the system through appropriate feedback and display of information. |
| Match | Match between system and world: The image of the system perceived by users should match the model the users have about the system. |
| Minimalist | Minimalist: Any extraneous information is a distraction and a slow-down. |
| Memory | Minimize memory load: Users should not be required to memorize a lot of information to carry out tasks. Memory load reduces users' capacity to carry out the main tasks. |
| Feedback | Informative feedback: Users should be given prompt and informative feedback about their actions. |
| Flexibility | Flexibility and efficiency: Users always learn and users are always different. Give users the flexibility of creating customization and shortcuts to accelerate their performance. |
| Message | Good error messages: The messages should be informative enough such that users can understand the nature of errors, learn from errors, and recover from errors. |
| Error | Prevent errors: It is always better to design interfaces that prevent errors from happening in the first place. |
| Closure | Clear closure: Every task has a beginning and an end. Users should be clearly notified about the completion of a task. |
| Undo | Reversible actions: Users should be allowed to recover from errors. Reversible actions also encourage exploratory learning. |
| Language | Use users' language: The language should be always presented in a form understandable by the intended users. |
| Control | Users in control: Do not give users that impression that they are controlled by the systems. |
| Document | Help and documentation: Always provide help when needed, ideally context-sensitive help. |

IV. RESULTS

*A. Questionnaires reliability*

The reliability of the heuristic questionnaire and NASA-TLX was measured using Cronbach's alpha. For the heuristic questionnaire, alpha was .98; for NASA-TLX, alpha was .79. As a rule of thumb, Leary [23] suggested that a Cronbach alpha in excess of .70 is generally adequate for newly developed questionnaires.

*B. Usability attributes*

A heuristic evaluation was conducted to identify usability issues faced by CT scan technicians. As shown in Fig. 2, the evaluation indicated a potentially catastrophic usability issue (i.e., leading to death) on all 14 tested usability attributes. The results in Fig. 3 show that technicians identified 529 issues in operating the CT scan, ranging in severity from Cosmetic to Catastrophe. Fig. 3 shows the combined severity for all 14 attributes, incorporating Cosmetic (88.2 cases), Minor (193.2 cases), Major (180.7 cases), and Catastrophe (66.9 cases).

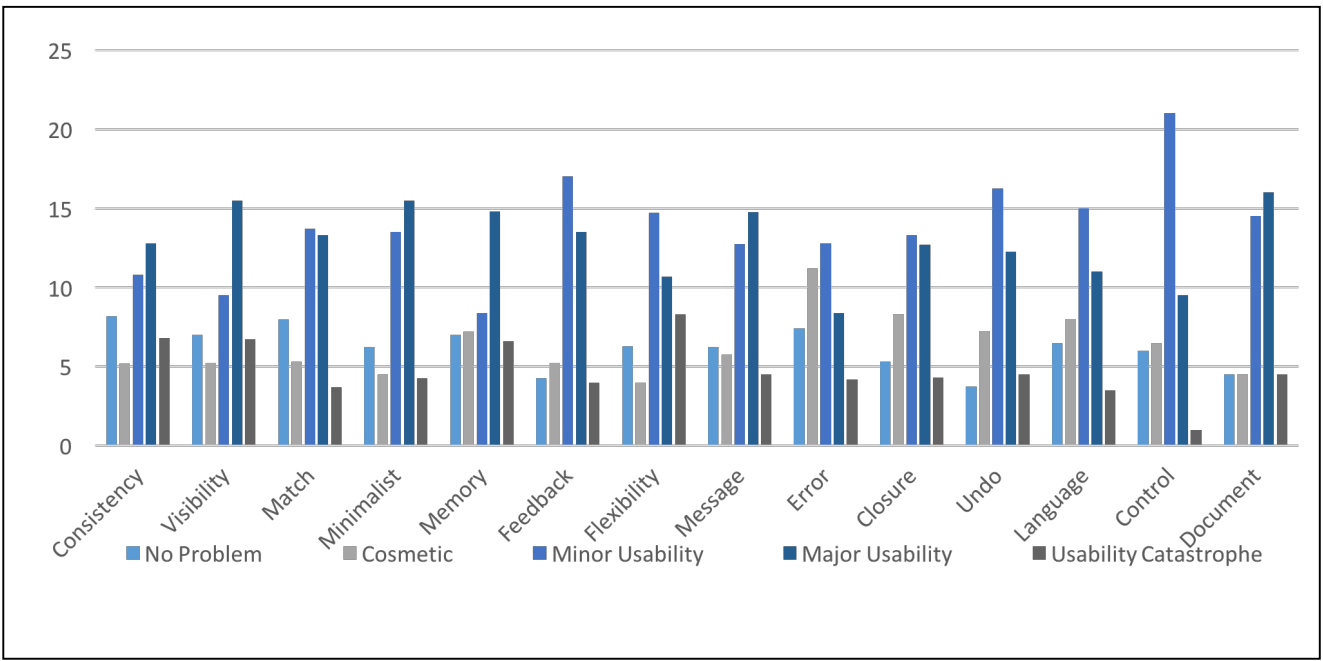

Fig. 2. Usability issues by attribute

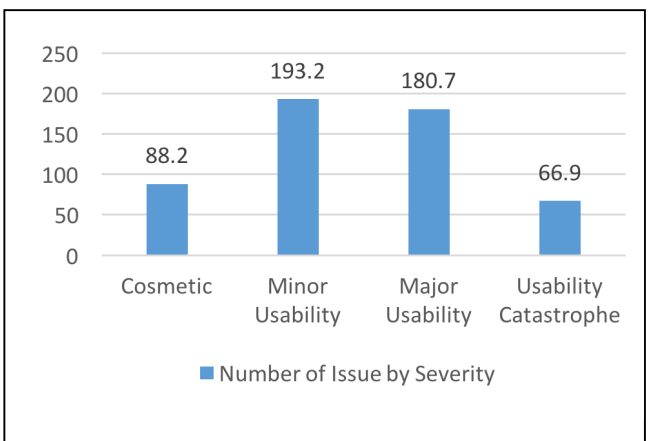

Fig. 3. Number of issues categorized by severity

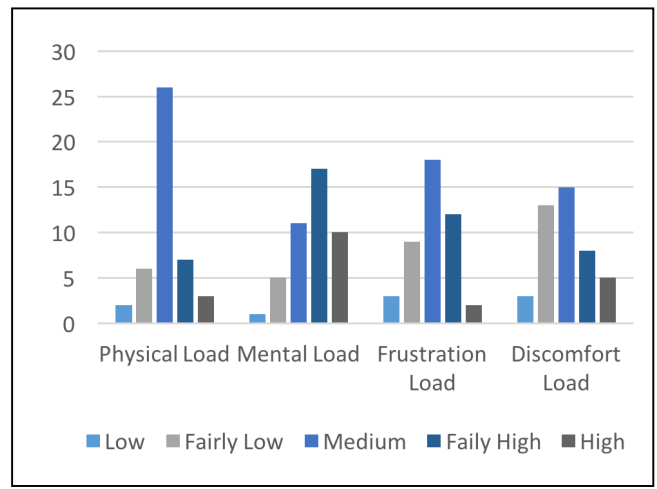

Fig. 4. NASA-TLX severity

*1) Gender*

Of 14 attributes, Mann Whitney tests showed a difference between Male and Female on Consistency (z = 2.21, p = .027, 2-sided); Flexibility (z = 1.99, p = .046, 2-sided); and Document (z = 2.09, p = .036, 2-sided).

*2) Age, Level of Education, and Years of Experience*

The nonparametric Kruskal-Wallis test showed no differences between category groups.

### C. NASA-TLX

NASA-TLX was used to identify the physical and mental loads associated with operating the CT scan. As shown in Fig. 4, a total of 3 cases registered high on physical load; 10 cases registered high on mental load; 2 cases registered high on frustration; and 5 cases registered high on discomfort.

*1) Gender*

A Mann Whitney test showed a difference between Male and Female on mental load (z = 3.23, p= .001, 2-sided).

*2) Age and Level of Education*

The nonparametric Kruskal-Wallis test showed no difference between category groups.

*3) Years of Experience*

A Kruskal-Wallis test showed a statistically significant difference by years of experience for at least one group on Frustration ($x^2$ = 10.9, p = .012) and Discomfort ($x^2$ = 9.2, p = .026). Dunn's pairwise test was performed for the six pairs of groups. There was strong evidence (p = .006, adjusted using the Bonferroni correction) of a difference in Frustration between 0–3 years and 8–11 years. The same pair also differed significantly (p = .036, adjusted using the Bonferroni correction) on Discomfort.

## V. Conclusion

Advances in technology entail advances in systems, increasing the pressure on users to manage complexity safely and efficiently. These findings support the mounting evidence of physical load (high = 3, fairly high = 7, medium = 26) where users see themselves as contributing physically to operate a CT. As 86.3% of users believed that operating CT scan involves medium to high mental load, manufacturers should pursue designs that reduce both physical and mental loads. Across the 14 usability attributes, 66.9 cases of Catastrophic usability were recorded. These should be fixed immediately before allowing the product to go to market. To ensure that users are willing and able to operate within the confines of rigid safety and regulatory guidelines, manufacturers should devote more effort to CT scan usability.

These findings add to mounting evidence that users differ according to gender and years of experience. In the context of CT scan operation, males and females different on the Consistency attribute (Sequence of action, Color, Layout, Font, Terminology, and Standards). For that reason, it is recommended that designers should provide customizable options to suit end-user needs and requirements. In addition, as males and females differed significantly on mental load, designers should take account of these differences in the design process and should actively iterate to the end of the process, testing and comparing in order to accommodate the usability needs of both genders.

## Acknowledgment

Saad Aldoihi thanks King Abdulaziz City Of Science and Technology for making this research possible by providing scholarship assistance. A thank you goes to Saudi Health Ministry for providing access to its hospitals and employees. A special thanks go to all participants who made the time and effort to participate in this study.